\documentclass[english,draftclsnofoot,12pt,onecolumn]{IEEEtran}
\usepackage[T1]{fontenc}
\usepackage[latin9]{inputenc}
\usepackage{float}
\usepackage{units}
\usepackage{amsmath}
\usepackage{amsthm}
\usepackage{amssymb}
\usepackage{stmaryrd}
\usepackage{graphicx}
\PassOptionsToPackage{normalem}{ulem}
\usepackage{ulem}

\makeatletter

\providecommand{\tabularnewline}{\\}
\floatstyle{ruled}
\newfloat{algorithm}{tbp}{loa}
\providecommand{\algorithmname}{Algorithm}
\floatname{algorithm}{\protect\algorithmname}

\theoremstyle{plain}
\newtheorem{thm}{\protect\theoremname}
\theoremstyle{plain}
\newtheorem{prop}[thm]{\protect\propositionname}

\usepackage{cite}
\usepackage{tikz,pgf}
\usetikzlibrary{calc,arrows}
\usepackage{algorithm}
\usepackage{algpseudocode}

\makeatother

\usepackage{babel}
\providecommand{\propositionname}{Proposition}
\providecommand{\theoremname}{Theorem}

\begin{document}
\title{Low-complexity Rank-Efficient Tensor Completion For Prediction And
Online Wireless Edge Caching\thanks{Authors are with Institute of digital communications, School of engineering,
The University of Edinburgh, Edinburgh, UK, EH9 3FG. Emails: \{ngarg,
t.ratnarajah\}@ed.ac.uk. }}
\author{Navneet Garg, \IEEEmembership{Member, IEEE}, and Tharmalingam Ratnarajah,
\IEEEmembership{Senior Member, IEEE}.}
\maketitle
\begin{abstract}
Wireless edge caching is a popular strategy to avoid backhaul congestion
in the next generation networks, where the content is cached in advance
at base stations to serve redundant requests during peak congestion
periods. In the edge caching data, the missing observations are inevitable
due to dynamic selective popularity. Among the completion methods,
the tensor-based models have been shown to be the most advantageous
for missing data imputation. Also, since the observations are correlated
across time, files, and base stations, in this paper, we formulate
the cooperative caching with recommendations as a fourth-order tensor
completion and prediction problem. Since the content library can be
large leading to a large dimension tensor, we modify the latent norm-based
Frank-Wolfe (FW) algorithm with towards a much lower time complexity
using multi-rank updates, rather than rank-1 updates in literature.
This significantly lower time computational overhead leads in developing
an online caching algorithm. With MovieLens dataset, simulations show
lower reconstruction errors for the proposed algorithm as compared
to that of the recent FW algorithm, albeit with lower computation
overhead. It is also demonstrated that the completed tensor improves
normalized cache hit rates for linear prediction schemes.
\end{abstract}

\section{Introduction}

With the continuous development of various intelligent devices and
various sized innovative application services such as high quality
video feeds, software updates, news updates, etc., wireless mobile
communications has been experiencing an unprecedented traffic surge
with a lot of redundant and repeated information, which limits the
capacity of the fronthaul and backhaul links \cite{8658196}. To lower
the redundant traffic, caching has emerged as an effective solution
for reducing the peak data rates by pre-fetching the most popular
contents in the local cache storage of the base stations (BS). In
the recent years, caching at the BS is actively feasible due to the
reduced cost and size of the memory \cite{li2018survey}. In \cite{li2018survey}
where cache-enabled networks are classified into macro-cell, heterogeneous
and D2D networks, given a set of a content library and the respective
content popularity  profile, content placement and delivery have been
investigated in order to optimize the backhaul latency delay in \cite{shanmugam2013femtocaching},
server load in \cite{poularakis2016complexity}, cache miss rate in
\cite{blaszczyszyn2015optimal,serbetci2017optimal,8490665}, etc.
With the known popularity profile, reinforcement learning approaches
\cite{sadeghi2018optimal,garg2020QFA} are studied for learning the
content placement. However, in practice, this profile is time-varying
and not known in advance, therefore, it needs to be estimated from
the past observations of the content requests. To estimate future
popularities, deep learning based prediction is employed with huge
training data in \cite{yin2018prediction,liu2018content}. In \cite{nakayama2015caching},
auto regressive (AR) prediction cache is used to predict the number
of requests in the time series. Linear prediction approach is investigated
for video segments in \cite{7990557}. To learn popularities independently
across contents, online policies are presented for cache-awareness
in \cite{8598719}, low complexity video caching in \cite{8658196,7488241},
user preference learning in \cite{8531745}, etc. These works on prediction
focus independently on the BSs to estimate the future content popularities.
However, the demands across files are correlated \cite{7931552},
since other similar contents can be served from the cache with the
same features as the requested content in order to maximize the cache
hit, which is also known as soft cache hit \cite{8374917}. Regarding
that in literature \cite{7931552,8352848,8510864,8704952,8725900,9410602,9457546},
recommendation based caching has been carried out based on low-rank
decomposition, deep reinforcement learning, etc. Moreover, the content
is also correlated across base stations, as studied in the recent
works on in-network caching solutions \cite{7996761,8424087}, where
base stations are jointly allocated cache contents. Furthermore, regarding
the correlation of popularities across time slots, in edge caching
literature, to avoid prediction, cache placement is performed to maximize
different objectives such as cache hit rate \cite{serbetci2017optimal},
average success probability \cite{8254810,8682841,8913496,garg2018online},
etc., given the past information up to the present time slot. The
maximization problems of these objectives can be simplified to the
prediction problem. Therefore, in this work, we focus on prediction
of such correlated demands to improve the caching performance using
a tensor approach. Due to these correlations and missing data, it
is difficult to store and predict the content popularities for large
content lirbary. Thus, after modeling the tensor completion problem,
we modify the Frank-Wolfe approach for the solution, followed by linear
prediction methods. A brief review of tensor completion approaches
is given as follows.

\subsection{Tensor completion methods}

In the past decades, tensor completion is intensively researched due
to its wide applications in a variety of fields, such as computer
vision \cite{8651300,8625383,romera2013new}, multi-relational link
prediction \cite{6979248,jenatton2012latent,guo2017efficient}, and
recommendation system \cite{doi:10.1002/widm.1201,ioannidis2019coupled}.
The goal of tensor completion is to recover an incomplete tensor from
partially observed entries. To the best of our knowledge, tensor completion
methods can be categorized into decomposition based and rank-minimization
based methods.

Decomposition based methods aim to factorize the incomplete tensor
into a sequence of low-rank factors and then predict the missing entries
via the latent factors. In recent years, CANDECOMP/PARAFAC (CP) decomposition
\cite{BRO1997149} and Tucker decomposition \cite{tucker1966some}
are the two most studied and popular models applied in tensor completion
in \cite{Acar_2011}, \cite{7010937}, \cite{6587455}. Although CP
and Tucker methods obtain good performance for low order tensors,
yet their performance rapidly degrades for higher-order tensors. Moreover,
the computing of CP-rank is NP hard and the number of parameters of
Tucker decomposition is exponential with the order of given tensors.
Recently, a tensor decomposition model, called Tensor-Ring (TR) decomposition
\cite{zhao2016tensor,zhao2019learning,9096509,8883205}, is proposed
to process high-order tensors. TR decomposition can express a higher-order
tensor by a multi-linear product over a sequence of lower-order latent
cores. A notable advantage of TR decomposition is that its total number
of parameters increases linearly with the order of the given tensor,
reducing the curse of dimensionality as compared to Tucker decomposition.
Attracted by these features, TR decomposition has drawn lots of attention
such as TR based weighted optimization \cite{8659708}, alternating
least squares \cite{8237869}, low-rank factors.

Rank-minimization based methods exploits the low-rank structure to
complete the tensor. Since the rank minimization is a non-convex and
NP-hard problem, overlapped nuclear norm \cite{6138863,7859390,Yu2019TensorringNN}
and latent nuclear norm \cite{NIPS2013_4985,8683685} have been defined
as the convex surrogates of tensor rank. The former norm assumes low-rank
across all modes and thus perform poorly, when the target is low-rank
in certain modes \cite{6138863}. The latter norm assumes few modes
in low-rank and often performs better than the former \cite{NIPS2013_4985}.
However, these two norms are based on the unbalanced mode-$k$ unfolding,
and thus, causing lack of capturing global information for higher-order
tensors.

\subsection{Contributions}

In this paper, inspired by the TR decomposition, we employ convex
TR-based latent nuclear norm \cite{9136661} based on cyclic unfolding,
i.e. overcoming the drawback of unbalanced unfolding. The tensor completion
task can be cast as a convex optimization problem to minimize the
latent norm, which is solved via modified Frank-Wolfe (FW) algorithm
\cite{9136661,pmlr-v28-jaggi13} with cyclic unfolding to provide
an efficient solution towards lower time complexity. In this work,
FW method is modified to obtain a low time-complexity solution via
gradient descent updates.  Furthermore, simulations on MovieLens
dataset are carried out to verify the algorithm and completion results
for cache hit rate in caching. The contributions of this paper can
be summarized as follows.

\subsubsection{Problem Formulation}

The missing entry problem in edge caching is cast as a tensor completion
problem, where the entries are correlated across base stations, files
and time. 

\subsubsection{Online solution}

To solve the tensor completion problem, we modify the FW algorithm
from \cite{9136661} towards a lower time complexity solution. The
modified approach is significantly faster than that in \cite{9136661}.
Using the proposed tensor completion, an online content caching algorithm
is presented to improve the cache hit rates. 

\subsubsection{Simulations and comparison}

Simulations are performed for MovieLens dataset towards the convergence
and observing the effect of latent factors. The cache hit rate performance
of caching is plotted for both mean based and linear prediction methods,
which also shows normalized cache hit rate improvements compared to
conventional CP decomposition (CPD) \cite{tensorlab3.0} and the FW
method in \cite{9136661}. For higher decomposition rank, the proposed
method provides better hit rates than that with \cite{9136661}.

\subsection*{Related work on TR based completion}

Related works include latent-norm based methods \cite{NIPS2013_4985,8683685,9136661}
and Tensor-Ring based methods \cite{8237869,Yu2019TensorringNN}.
\cite{NIPS2013_4985} employed the latent nuclear norm, while \cite{8683685}
defined a new latent nuclear norm via Tensor Train. However, they
are based on unbalanced mode-$k$ unfolding. \cite{8237869} applied
TR decomposition with alternating least squares for completion, while
\cite{yuan2018tensor} proposed TR low-rank factors. To reduce the
computational complexity per iteration, \cite{Yu2019TensorringNN}
utilize an overlapped TR nuclear norm. Further, to reduce the number
of parameters for selection, \cite{9136661} proposed a new latent
TR-nuclear norm.

\subsection*{Organization}

The rest of this paper is organized as follows. The prediction in
the edge caching framework is presented in section \ref{sec:System-model}.
In Section \ref{sec:Tensor-Completion-Algorithm}, tensor completion
algorithm is provided. Section \ref{sec:Simulation-Results} investigates
simulation results for the real-world dataset. Finally, the paper
is concluded in section \ref{sec:Conclusion}.

\subsection*{Notations}

Scalars, vectors, and matrices are respectively denoted by lowercase,
boldface lowercase, and bold capital letters. A tensor of order $N>3$
is denoted by calligraphic letter $\mathcal{X}$. The notation $\mathcal{X}(i_{1},i_{2},\ldots,i_{N})$
represents an element in $X$, while $\mathcal{X}(:,i_{2},\ldots,i_{N})$
and $\mathcal{X}(:,:,i_{3},\ldots,i_{N})$ denotes a fiber along mode
$1$ and a slice along mode 1 and mode 2 respectively. Inner product
of two tensors $\mathcal{X}$ and $\mathcal{Y}$ of the same size
is given as $\left\langle \mathcal{X},\mathcal{Y}\right\rangle =\sum_{i_{1},i_{2},\ldots,i_{N}}\mathcal{X}\left(i_{1},i_{2},\ldots,i_{N}\right)\mathcal{Y}\left(i_{1},i_{2},\ldots,i_{N}\right)$
and the Frobenius norm can be obtained as $\left\Vert \mathcal{X}\right\Vert _{F}^{2}=\left\langle \mathcal{X},\mathcal{X}\right\rangle $.
Notations $\text{tr}(\mathbf{A})$ and $\|\mathbf{A}\|_{*}$ defines
the trace and the nuclear norm of a matrix $\mathbf{A}$. 

\begin{table}
\centering %
\begin{tabular}{|l|l|}
\hline 
Symbol & Description\tabularnewline
\hline 
\hline 
$N_{BS}$ & Number of base stations\tabularnewline
\hline 
$\mathcal{F}$, $F$ & Content library and its size\tabularnewline
\hline 
$L_{BS}=\left|\mathcal{C}_{bt}\right|$ & Cache size at a BS\tabularnewline
\hline 
$\mathbf{c}_{bt},\mathbf{c}_{bt}(f)$ & Cache status and $f^{th}$ file status\tabularnewline
\hline 
$b$ & BS index\tabularnewline
\hline 
$t$ & Time slot index\tabularnewline
\hline 
$f$ & Content library index\tabularnewline
\hline 
$\mathcal{D}_{t}$,$D_{fibj}$ & Observed tensor of \#requests, and its entries\tabularnewline
\hline 
$\tau$ & Number of time slots for prediction\tabularnewline
\hline 
$\mathcal{H}_{b,t}$ & Cache hit rate \tabularnewline
\hline 
$\bar{D}_{fbt}$,$\hat{\bar{D}}_{fbt}$ & Normalized and estimated \#requests\tabularnewline
\hline 
$M$,$c_{bi}$ & Order and coefficients of linear prediction\tabularnewline
\hline 
$d$ & For cyclic unfolding/folding\tabularnewline
\hline 
$N$ & Number of tensor dimensions\tabularnewline
\hline 
$\mathcal{I}$ & Binary (0 or 1) valued tensor\tabularnewline
\hline 
$\mathcal{T}$ & Observed tensor for completion\tabularnewline
\hline 
$\mathcal{X}$ & Tensor to be determined\tabularnewline
\hline 
$\mathcal{X}_{k}$ & $k^{th}$ component s.t. $\mathcal{X}=\sum_{k=1}^{N}\mathcal{X}_{k}$\tabularnewline
\hline 
$\mathcal{X}_{k,(k,d)}$ & Cyclic unfolding of $\mathcal{X}_{k}$ of size $\bar{I}_{k}\times\bar{J}_{k}$\tabularnewline
\hline 
$\mathcal{S}$ & Gradient representation\tabularnewline
\hline 
$\gamma$,$\beta$ & Step size, norm constraint\tabularnewline
\hline 
$R_{k}$, $R$ & $k^{th}$ mode rank and rank constraint\tabularnewline
\hline 
$r_{k}$ & Rank of current SVD\tabularnewline
\hline 
$\mathbf{U}_{k},\mathbf{V}_{k},\Sigma_{k}$ & Decomposition of $\mathcal{X}_{k}$\tabularnewline
\hline 
\end{tabular}

\caption{List of variables.\label{tab:List-of-variables.}}

\end{table}

\section{System model\label{sec:System-model}}

\begin{figure}
\centering\includegraphics[width=1\columnwidth]{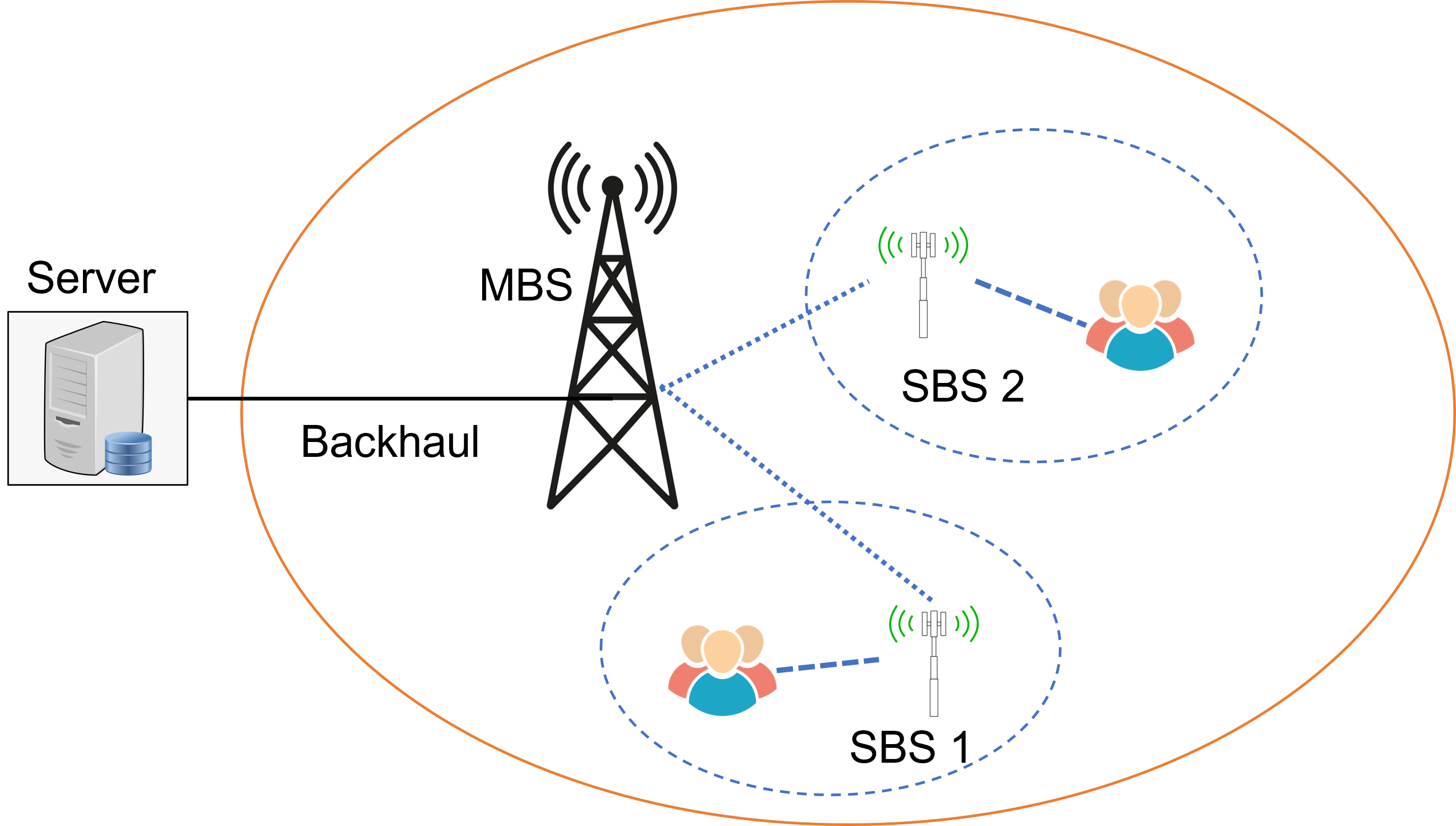}

\caption{System model of the caching framework. \label{fig:System-model-of}}
\end{figure}

We consider a multi-cell network with one macro base station (MBS)
and $N_{BS}$ small base stations (SBS), where each SBS serves multiple
users. An example of this system is illustrated in Figure \ref{fig:System-model-of}.
 Each user requests contents from a fixed library. Let the content
library be indexed by the set $\mathcal{F}=\left\{ 1,\ldots,F\right\} $,
where each content is assumed to be of equal size .

In the time slot $t\in\left\{ 1,\ldots,T\right\} $, the $b^{th}$
SBS has a cache of size $L_{BS}$, and the $F\times1$ vector $\mathbf{c}_{bt}$
describes the status of cache, that is, if the $f^{th}$ content is
cached, $\mathbf{c}_{bt}(f)=1$ (else $0$ for not cached) with the
cache size constraint $\mathbf{c}_{bt}^{T}\mathbf{1}_{F}=L_{BS}$.
For fractional caching, where a portion of content is cached rather
than the whole file, we have $\mathbf{c}_{bt}(f)\in\left[0,1\right]$
denoting the fraction of the $f^{th}$ content being cached, for each
$f\in\mathcal{F}$. Before delivering the requested contents from
users, a subset of popular contents are cached in the cache of the
$b^{th}$ base station. Users are provided with the recommendations
of the contents from library in a decreasing order of popularities
at the local SBS. Based on the requested contents from the library,
we define a direct hit to be the cache hit when the requested content
is present in the cache, whereas indirect hits are the hits in cache
for which users choose as an alternative to the requested content
based on the recommendations when the requested content is unavailable
in the cache. Each SBS collects the data about the direct and indirect
hits for each time slot.

Let $\mathcal{D}_{t}=\big\{ D_{fibj}\in\mathbb{R}_{+},\forall f,i\in\mathcal{F},b=1,\ldots,N_{BS},j=t-\tau+1,\ldots,t\big\}$
be a fourth-order tensor representing the aggregated data of number
of direct and indirect content requests across all SBSs and for previous
$\tau$ time slots ($t-\tau+1,\ldots,t$). The four dimensions of
the tensor $\mathcal{D}_{t}$ respectively represent the requested
content's index, indices of requested contents based on recommendations,
base station index, and time slot. In other words, in the $t^{th}$
time slot at $b^{th}$ BS, the number of recommended requests for
the $i^{th}$ file, when the $f^{th}$ content is primarily requested,
is denoted as $D_{fibt}$. Thus, the size of tensor is $I_{1}\times I_{2}\times I_{3}\times I_{4}$,
where $I_{1}=I_{2}=F,\,I_{3}=N_{BS}$ and $I_{4}=\tau$. Note that
only few entries of the tensor $\mathcal{D}_{t}$ can be observed,
since the library is large and a few files are popular in a given
time slot. Therefore, before processing this sparse tensor to derive
the cache placement scheme, a tensor completion approach is essential.
 The following subsection describes the dynamics of the caching system.

\subsection{Edge caching procedures}

\begin{figure}
\centering \includegraphics[width=8.9cm]{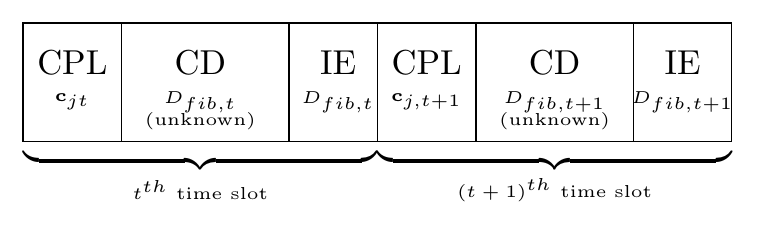}

\caption{Time slot structure. CPL: content placement, CD: content delivery,
IE: information exchange. \label{fig:Time-slot-structure.}}

\end{figure}

The structure of the time slot is shown in Figure \ref{fig:Time-slot-structure.}.
 The first phase is the content placement (CPL) phase, where contents
are placed in each BS's cache based on the present information at
base stations. The subsequent phase is the content delivery phase,
where the content is delivered from the cache as per the requests
from users. The next phase is dedicated for information exchange,
where multiple base stations exchange the information about the number
of hits and requests for better cache placements in the next time
slot. For the CPL phase, the content placement strategy is chosen
as a combination of two methods, that is, linear prediction of contents'
demands followed by most-popular caching (MPC) scheme. Since accurate
prediction requires the data available for different contents, tensor
completion problem is considered prior to performing prediction.

Moreover, in practice, the demands (number of requests for contents)
are correlated across different contents and base stations, reducing
the rank of the tensor $\mathcal{D}$. Thus, an independent prediction
for individual base station and for each file can cause performance
degradation. To deal with these correlation issues, authors in literature
considers in-network caching \cite{9053906}, joint reinforcement
learning \cite{sadeghi2018optimal}, etc. However, in this work, we
focus on improving the caching performance using tensor completion
methods. 

\subsection{Normalized cache hit rate and content placement}

In literature \cite{8913496,8531745}, to measure caching performance,
several measures like cache hit (miss) rate, average success probability
(ASP), etc have been considered. Improvements in these measures also
relies on the prediction of demands. Cache hit rate at the $b^{th}$
base station in the $t^{th}$ time slot can be written as 
\begin{equation}
\mathcal{H}_{bt}=\frac{\sum_{f,i\in\mathcal{F}}D_{fibt}\mathbf{c}_{bt}(f)}{\sum_{f,i\in\mathcal{F}}D_{fibt}},
\end{equation}
which evaluates the cache placement policy. 

Note that the number of requests at  SBSs is random with unknown distribution,
and not available in advance. That is, the decision for the $f^{th}$
file is set based on the previous number of requests. The objective
of caching is to find the best content placement strategy to maximize
the hit rate above. If in $(t+1)^{th}$ time slot the number of requests
$D_{fib,t+1}$ is known in advance, we can choose $L_{BS}$ files
with largest number of requests. On the other hand, when $D_{fib,t+1}$
is not known, it is natural to maximize the expected value given the
previous information as
\begin{subequations}
\begin{align}
 & \arg\max_{\mathbf{c}_{b,t+1},\forall b}\mathbb{E}_{D}\left[\sum_{b}\mathcal{H}_{b,t+1}|\mathcal{D}_{t}\right]\\
 & =\arg\max_{\mathbf{c}_{b,t+1},\forall b}\sum_{b}\sum_{f\in\mathcal{F}}\mathbf{c}_{bt}(f)\mathbb{E}_{D}\left[\frac{\sum_{i\in\mathcal{F}}D_{ifb,t+1}}{\sum_{f,i\in\mathcal{F}}D_{ifb,t+1}}|\mathcal{D}_{t}\right],
\end{align}
\end{subequations}
in which $\mathbb{E}_{D}\left[\frac{\sum_{i\in\mathcal{F}}D_{ifb,t+1}}{\sum_{f,i\in\mathcal{F}}D_{ifb,t+1}}|\mathcal{D}_{t}\right]$
denotes the conditional mean estimate of the normalized number of
requests. This estimate is the prediction of demands at $t+1$, given
the past $\tau$ observations of demands until time slot $t$. Note
that given the prediction estimates, the solution for caching strategy
includes the $L_{BS}$ contents which have largest values in the estimate.
 To compute this estimate, the probability distribution of normalized
requests must be known. However, the number of users' requests are
random and non-stationary with unknown distribution. Therefore, in
the following, we obtain linear prediction estimation using least
squares. 

\subsection{Linear prediction}

Let $\bar{D}_{fbt}=\frac{\sum_{i\in\mathcal{F}}D_{ifbt}}{\sum_{f,i\in\mathcal{F}}D_{ifbt}}$
denote the normalized number of requests, such that $\sum_{f\in\mathcal{F}}\bar{D}_{fbt}=1$.
To obtain the linear predicton, the normalized demands are assumed
to evolve as a linear combination of demands in the temporal dimension
as 
\begin{equation}
\bar{D}_{fb,t+1}\approx\sum_{m=1}^{M}c_{bm}\bar{D}_{fbm},\forall f\in\mathcal{F},
\end{equation}
where $M$ is the order of prediction and $c_{bm}$ are the prediction
coefficients. These coefficients are obtained by solving least squares
fit problem, given the previous $\tau$ observations $\bar{D}_{fbj},j=t-\tau+1,\ldots,t$
for each $b$ as 
\begin{subequations}
\begin{align}
\min_{c_{bi}\forall i} & \sum_{f\in\mathcal{F}}\sum_{j=0}^{\tau-M-1}\left|\bar{D}_{fb,t-j}-\sum_{m=1}^{M}\bar{D}_{fb,t-j-m}c_{bm}\right|^{2}\\
\text{subject to} & \sum_{m=1}^{M}c_{bm}\bar{D}_{fbm}\geq0,\forall f\in\mathcal{F}\\
 & \sum_{f\in\mathcal{F}}\sum_{m=1}^{M}c_{bm}\bar{D}_{fbm}=1,
\end{align}
\end{subequations}
where the constraints provides the non-negative values. Let the prediction
estimate be denoted as $\hat{\bar{D}}_{fb,t+1}=\sum_{m=1}^{M}c_{bm}\bar{D}_{fbm}$.
Mean based approach can be considered as a special case of linear
prediction approach, i.e. $c_{bi}=\nicefrac{1}{d},\forall i=1,\ldots,d$.
Since the data is sparse and correlated across files and base stations,
it is difficult to directly obtain accurate predictions via these
methods. Therefore, after filling the entries via tensor completion
method, the above linear prediction can be used to find the future
popularity estimates. For simplicity, in later sections, we recall
this linear prediction method using the notation $\left\{ \hat{\bar{D}}_{fb,t+1},\forall f,b\right\} \leftarrow LP\left\{ \mathcal{D}_{t}\right\} $.

\paragraph*{}

\subsection{Tensor preliminaries: Circular unfolding and folding}

To efficiently represent the information in higher order tensors,
authors in \cite{Yu2019TensorringNN,8659492} defined a balance unfolding
scheme based on circular unfolding. For an $N$-order tensor $\mathcal{X}$,
the tensor circular unfolding matrix, denoted by $\mathcal{X}_{(k,d)}$
of size $\bar{I}_{k}\times\bar{J}_{k}=I_{a}I_{a+1}\ldots I_{k}\times I_{k+1}\ldots I_{a-1}$,
can be written as 
\begin{equation}
\mathcal{X}_{\left(k,d\right)}\left(i_{a}i_{a+1}\ldots i_{k}\times i_{k+1}\ldots i_{a-1}\right)=\mathcal{X}\left(i_{1},i_{2},\ldots,i_{N}\right),
\end{equation}
where $d$ is a positive integer and 
\begin{equation}
a=\begin{cases}
k-d+1, & d\leq k;\\
k-d+1+N, & d>k.
\end{cases}
\end{equation}
The above unfolding with the given mode $k$ and shift $d$ provides
the balanced unfolding.  Note that the balance of the above unfolding
depends on the shift $d$. Similarly, the folding of a matrix $\mathbf{X}$
of size $I_{a}I_{a+1}\ldots I_{k}\times I_{k+1}\ldots I_{a-1}$ along
the mode $k$ and shift $d$ provides the tensor $\left\llbracket \mathbf{X}\right\rrbracket _{\left(k,d\right)}$
of size $I_{1}\times\ldots\times I_{N}$.

\subsection{Latent nuclear norm with cyclic unfolding}

The cyclic unfolding based latent nuclear norm is defined for an $N$-order
tensor $\mathcal{X}$ as follows \cite{guo2017efficient,9136661}
\begin{equation}
\left\Vert \mathcal{X}\right\Vert _{TR}=\min_{\mathcal{X}_{1}+\ldots+\mathcal{X}_{N}=\mathcal{X}}\sum_{k=1}^{N}\left\Vert \mathcal{X}_{k,(k,d)}\right\Vert _{*},\label{eq:TRnorm}
\end{equation}
where the minimum is over $N$ tensors $\left\{ \mathcal{X}_{k}\right\} _{k=1}^{N}$,
and $\mathcal{X}_{k,(k,d)}$ denotes the low-rank unfolding of $\mathcal{X}_{k}$
along the mode $k$ with a given value of $d$. 

\subsection{Tensor completion problem}

Let $\mathcal{T}\in\mathbb{R}^{I_{1}\times\cdots\times I_{N}}$ be
an $N$-order sparse tensor with the observed entries. The location
of entries in $\mathcal{T}$ is denoted by another indicator tensor
$\mathcal{I}$, where $\mathcal{I}\left(i_{1},i_{2},\ldots,i_{N}\right)$
is 1 when $\mathcal{T}\left(i_{1},i_{2},\ldots,i_{N}\right)\neq0$,
and 0, otherwise. The notation $\left|\mathcal{I}\right|$ and $\mathcal{T}(\mathcal{I})$
define the number of non-zeros in the $\mathcal{I}$ and the entries
of $\mathcal{T}$ for the corresponding non-zeros indices in $\mathcal{I}$,
respectively.

The optimization problem for low-rank tensor completion using TR-latent
nuclear norm is cast as 
\begin{subequations}
\begin{align}
\min_{\mathcal{X}_{k},\forall k} & \left\Vert \mathcal{X}\right\Vert _{TR}\label{eq:OPT-1}\\
\text{subject to } & \mathcal{X}=\sum_{k=1}^{N}\mathcal{X}_{k},\mathcal{X}(\mathcal{I})=\mathcal{T}(\mathcal{I}),
\end{align}
\end{subequations}
where $\mathcal{X}_{k}$ are component tensors and unfolded cyclically.
To solve the above optimization, recently in \cite{9136661,guo2017efficient},
Frank Wolfe algorithm is used to obtain parameters independent iterative
procedure for tensor completion. However, the computation complexity
is large due to optimum mode selection $k$, which requires SVD of
$\mathcal{X}_{k}$ along each of $N$ modes (since $k=1,\ldots,N$);
and the iterative compression of basis matrices, where SVD, QR and
a quadratic optimization is performed. In this work, we propose a
procedure with much lower computational overhead and along with significantly
better reconstruction error. The details are described in the following.

\section{Tensor completion algorithm\label{sec:Tensor-Completion-Algorithm}}

\subsection{Rank efficient modified FW algorithm}

Under the Frank-Wolfe framework, the optimization problem in \eqref{eq:OPT-1}
can be rewritten as 
\begin{subequations}
\begin{align}
\min_{\mathcal{X}} & F(\mathcal{X})\\
\text{subject to} & \left\Vert \mathcal{X}\right\Vert _{TR}\leq\beta,\label{eq:FW-con}
\end{align}
\end{subequations}
where $F(\mathcal{X})=\frac{1}{2}\left\Vert \mathcal{X}(\mathcal{I})-\mathcal{T}(\mathcal{I})\right\Vert _{F}^{2}$
and $\beta>0$. The constraint $\left\Vert \mathcal{X}\right\Vert _{TR}\leq\beta$
is also related to the rank constraint, that is, if the rank constraint
is not specified, the rank of the solution can be high. However, for
the proposed procedure, we will show that with the rank constraint
specified, the above constraint can be relaxed; in other words, low
rank solutions can be obtained with the specified $\beta$ limit. 

The above optimization is solved via the gradient descent steps in
the FW algorithm. It can be observed that the constraint $\mathcal{X}=\sum_{k=1}^{N}\mathcal{X}_{k}$
is also included in the TR-norm constraint. For the cyclic unfolding,
we can write 
\begin{equation}
\mathcal{X}=\sum_{k=1}^{N}\left\llbracket \mathcal{X}_{k,\left(k,d\right)}\right\rrbracket _{\left(k,d\right)},
\end{equation}
where the matrices $\mathcal{X}_{k,\left(k,d\right)}$ are of low-rank,
say $R_{k}$. Given the rank constraint of overall TC problem (say
$R$), all $\mathcal{X}_{k}$ must satisfy $\sum_{k=1}^{N}R_{k}\leq R$.
Note that for notational simplicity, we omit the iteration index.
The update of the gradient descent can be given as 
\begin{equation}
\mathcal{X}\leftarrow\mathcal{X}-\gamma\mathcal{S},\label{eq:X_update}
\end{equation}
where $\gamma>0$ and the tensor $\mathcal{S}$ represent the gradient
of $F(\mathcal{X})$, $\nabla F=\mathcal{X}(\mathcal{I})-\mathcal{T}(\mathcal{I}).$
The efficient representation of $F(\mathcal{X})$ with a decomposition
and the TR-norm constraint can be obtained as follows.

\subsubsection{Constraint linear optimization}

The problem of finding the tensor gradient of $F(\mathcal{X})$ to
satisfy the constraint \eqref{eq:FW-con} can be expressed as 
\begin{align*}
\mathcal{S} & =\arg\max_{\left\Vert \mathcal{S}\right\Vert _{TR}\leq\beta}\left\langle \mathcal{S},\nabla F\right\rangle ,
\end{align*}
where the objective is maximize the correlation between the above
two tensors. Note that the objective function is linear. One possible
solution is to choose $\mathcal{S}=\beta\frac{\nabla F}{\left\Vert \nabla F\right\Vert _{TR}}$.
However, we also need low-rank decomposition of $\mathcal{S}$ for
the tensor completion problem. Therefore, we first find the optimum
mode for cyclic unfolding, and then, leverage SVD for decomposition
components. To find the optimum mode, we compare the first dominant
singular value of the unfolded tensor $\nabla F$ along different
modes, that is, 
\[
k^{*}=\arg\max_{k}\sigma_{max}\left[\left(\nabla F\right)_{\left(k,d\right)}\right],
\]
where the notation $\sigma_{max}(\mathbf{A})$ denotes the maximum
eigenvalue of the matrix $\mathbf{A}$. Let the folded gradient has
SVD as $\left(\nabla F\right)_{\left(k,d\right)}=\tilde{\mathbf{U}}_{k}\tilde{\Sigma}_{k}\tilde{\mathbf{V}}_{k}^{T}$,
where the singular values are assumed in a descending order. For a
low-rank tensor completion, rank is a constraint, say $r_{k}$. We
will present how to obtain $r_{k}$ later. Therefore, we have the
gradient solution as 
\begin{equation}
\mathcal{S}=\beta\frac{\left\llbracket \tilde{\mathbf{U}}_{k^{*}}(1:r_{k^{*}})\tilde{\Sigma}_{k^{*}}(1:r_{k^{*}},1:r_{k^{*}})\tilde{\mathbf{V}}_{k^{*}}^{T}(1:r_{k^{*}})\right\rrbracket _{\left(k^{*},d\right)}}{\mathbf{1}_{r_{k^{*}}}^{T}\tilde{\Sigma}_{k^{*}}(1:r_{k^{*}},1:r_{k^{*}})\mathbf{1}_{r_{k^{*}}}},\label{eq:S_eqn}
\end{equation}
where for a matrix $\mathbf{A}(1:m,1:n)$ denotes the submatrix with
entries belonging to the first $m$ rows and first $n$ columns of
$\mathbf{A}$; and $\mathbf{A}(1:n)$ is the submatrix with first
$n$ columns of $\mathbf{A}$; $\mathbf{1}_{n}$ denotes $n\times1$
vectors of ones. 

\emph{Remark (rank-1 updates)}: For gradient representation $\mathcal{S}$,
instead of choosing $r_{k}$ rank, rank-$1$ updates can be considered
(for a suboptimal solution with larger overhead) as 
\begin{equation}
\mathcal{S}=\beta\left\llbracket \tilde{\mathbf{U}}_{k^{*}}(1)\tilde{\mathbf{V}}_{k^{*}}^{T}(1)\right\rrbracket _{\left(k^{*},d\right)},
\end{equation}
which is inefficient than as compared to \eqref{eq:S_eqn} due to
the fact that the structure of singular values $\tilde{\Sigma}_{k^{*}}$
is absent. Algorithm in \cite{9136661,guo2017efficient} gathers many
such singular vector (and values), and find the structure of singular
values via the compression step.

\emph{Remark (max-mode selection)}: In the above step, the mode $k^{*}$
is selected based on maximum singular value. For an $I_{1}\times\cdots\times I_{N}$
tensor, the singular value of an unfolding is maximum when its dimensions
are minimum. If the $k^{th}$ unfolding has dimension $\bar{I}_{k}\times\bar{J}_{k}$,
then the value $k^{*}$ can be selected as 
\begin{equation}
k^{*}=\arg\min_{k}\min\left\{ \bar{I}_{k},\bar{J}_{k}\right\} ,
\end{equation}
which can significantly reduce the computational complexity of mode-selection,
that is, the computations for $N$ singular values.

\subsubsection{Line search}

The line search problem is to find the step size, which can be expressed
as 
\begin{subequations}
\begin{align}
\gamma & =\arg\min_{\gamma\geq0}\left\Vert \left\{ \mathcal{X}\left(\mathcal{I}\right)-\gamma\mathcal{S}\right\} -\mathcal{T}\left(\mathcal{I}\right)\right\Vert _{F}^{2}\\
 & =\arg\min_{\gamma\geq0}\bar{a}\gamma^{2}-2\bar{b}\gamma\\
 & =\max\left\{ \frac{\bar{b}}{\bar{a}},0\right\} ,\label{eq:step_size}
\end{align}
\end{subequations}
where $\bar{a}=\left\Vert \mathcal{S}\left(\mathcal{I}\right)\right\Vert _{F}^{2}$,
$\bar{b}=\left\langle \mathcal{X}\left(\mathcal{I}\right)-\mathcal{T}\left(\mathcal{I}\right),\mathcal{S}\left(\mathcal{I}\right)\right\rangle $,
and the solution is obtained by differentiation. The max-operator
arises due to $\gamma\geq0$ constraint. With $\gamma$ and $\mathcal{S}$
obtained, the tensor $\mathcal{X}$ can be updated using the equation
\eqref{eq:X_update}. 

\subsubsection{Decomposition update}

Thanks to the decomposition, one does not need to store the whole
tensor $\mathcal{S}$ or $\mathcal{X}$. We can store SVD components
and step size as a representation of the update of the unfolded-component
tensors. In other words, for the selected optimum mode-$k^{*}$, singular
vectors along each mode are updated from, the update equation \eqref{eq:X_update}
can be simplified as 
\begin{subequations}
\begin{align}
 & \mathcal{X}\leftarrow\mathcal{X}-\gamma\mathcal{S}\\
 & =\sum_{k=1}^{N}\left\llbracket \mathcal{X}_{k,\left(k,d\right)}\right\rrbracket _{\left(k,d\right)}-\gamma\mathcal{S}\\
 & =\sum_{k\neq k^{*}}\left\llbracket \mathcal{X}_{k,\left(k,d\right)}\right\rrbracket _{\left(k,d\right)}+\Bigg\llbracket\mathcal{X}_{k^{*},\left(k^{*},d\right)}-\\
 & \gamma\beta\frac{\tilde{\mathbf{U}}_{k^{*}}(1:r_{k^{*}})\tilde{\Sigma}_{k^{*}}(1:r_{k^{*}},1:r_{k^{*}})\tilde{\mathbf{V}}_{k^{*}}^{T}(1:r_{k^{*}})}{\mathbf{1}_{r_{k^{*}}}^{T}\tilde{\Sigma}_{k^{*}}(1:r_{k^{*}},1:r_{k^{*}})\mathbf{1}_{r_{k^{*}}}}\Bigg\rrbracket_{\left(k^{*},d\right)}.\nonumber 
\end{align}
\end{subequations}
Since for each $k$, we have $\mathcal{X}_{k,\left(k,d\right)}=\mathbf{U}_{k}\boldsymbol{\Sigma}_{k}\mathbf{V}_{k}^{T}$,
the above equation leads to the updation of singular vectors along
only $k^{*}$-th mode as 
\begin{subequations}
\begin{align}
\mathbf{U}_{k^{*}} & \leftarrow\left[\mathbf{U}_{k^{*}},-\tilde{\mathbf{U}}_{k^{*}}(1:r_{k^{*}})\right],\\
\mathbf{V}_{k^{*}} & \leftarrow\left[\mathbf{V}_{k^{*}},\tilde{\mathbf{V}}_{k^{*}}(1:r_{k^{*}})\right],\\
\boldsymbol{\Sigma}_{k^{*}} & \leftarrow\left[\begin{array}{cc}
\boldsymbol{\Sigma}_{k^{*}} & \mathbf{0}\\
\mathbf{0} & \frac{\gamma\beta\tilde{\Sigma}_{k^{*}}(1:r_{k^{*}},1:r_{k^{*}})}{\mathbf{1}_{r_{k^{*}}}^{T}\tilde{\Sigma}_{k^{*}}(1:r_{k^{*}},1:r_{k^{*}})\mathbf{1}_{r_{k^{*}}}}
\end{array}\right],\\
R_{k^{*}} & \leftarrow R_{k^{*}}+r_{k^{*}},
\end{align}
\end{subequations}
where the components $\mathbf{U}_{k},\boldsymbol{\Sigma}_{k},\mathbf{V}_{k},R_{k},k\neq k^{*}$
remains unchanged during this iteration. Regarding the update of $r_{k}$,
there are two constraints, rank of the unfolded matrix $r_{k}\leq\min\left\{ \bar{I}_{k},\bar{J}_{k}\right\} -R_{k}$,
and the overall rank constraint $r_{k}\leq R-\sum_{k=1}^{N}R_{k}$.
Thus, for the next iteration, the update for $r_{k}$ can be written
as 
\begin{equation}
r_{k}\leftarrow\min\left\{ \bar{I}_{k}-R_{k},\bar{J}_{k}-R_{k},R-\sum_{i=1}^{N}R_{i}\right\} ,
\end{equation}
which also defines the stopping criteria, that is, the iterative procedure
stops if $r_{k}$ reaches $0$. 

\emph{Remark (Compression step)}: It can be seen that due to rank
constraint $r_{k}$, the above representation does not explode into
many basis matrices. Thus, no-compression is required here, which
significantly reduces the computational overhead, as compared to \cite{9136661,guo2017efficient}.
 Further, if the compression step is leveraged into the proposed
algorithm, the reconstruction errors can be further reduced. However,
for an online algorithm, this step is omitted. 

\subsubsection{Algorithm}

The Algorithm \ref{alg:ALGO1} presents the proposed TC procedure,
which combines the steps obtained in the above subsections. In addition
to the input observed tensor $\mathcal{T}$, the method requires to
specify the required rank $R$ and the low-rank reconstruction error
limit $\beta$. After the initialization of variables $\mathcal{X}=0$,
$R_{k}=r_{k}=0,\mathbf{U}_{k}=\Sigma_{k}=\mathbf{V}_{k}=\emptyset,\forall k$,
first the optimum mode $k^{*}$ is selected and the corresponding
unfolded mode SVD of $\nabla F$ is computed. Based on the value of
$r_{k}$, the gradient representation $\mathcal{S}$ and the step
size $\gamma$ are calculated via rank-$r_{k}$ truncated SVD. Subsequently,
the rest of variables are updated including $\mathcal{X}$, $\mathbf{U}_{k^{*}},\boldsymbol{\Sigma}_{k^{*}},\mathbf{V}_{k^{*}}$
and $R_{k^{*}}$. For usage, the algorithmic procedure is denoted
as $\mathcal{X}\leftarrow TCA(\mathcal{T},R)$. 

The value of $r_{k}$ denotes the number of available dimensions in
the $k^{th}$ mode. The value $r_{k}=0$ means either, the overall
rank constraint $R=\sum_{k}R_{k}$ is satisfied, or, mode-$k$ rank
is reached, i.e., $R_{k}=\min\left\{ \bar{I}_{k},\bar{J}_{k}\right\} $.
The former constraint leads to the stopping criteria, while the latter
adds the pruning step for the search space of optimum mode selection,
that is, $\mathcal{N}\leftarrow\mathcal{N}\setminus\left\{ k\right\} $.
\begin{prop}
\label{prop:The-propositionbeta}Given the rank constraint $R$, the
Algorithm \ref{alg:ALGO1} is independent of $\beta$. 
\end{prop}
\begin{IEEEproof}
In the algorithm, the update equation \eqref{eq:X_update} depends
on the the product $\gamma\beta$. By the definition of $\gamma$,
we write $\gamma=\frac{\left\langle \mathcal{X}\left(\mathcal{I}\right)-\mathcal{T}\left(\mathcal{I}\right),\mathcal{S}\left(\mathcal{I}\right)\right\rangle }{\left\Vert \mathcal{S}\left(\mathcal{I}\right)\right\Vert _{F}^{2}}.$
Substituting the value $\mathcal{S}$ from \eqref{eq:S_eqn} provides
$\beta\gamma=constant$, where the constant is specified via the singular
values in the unfolding in $k^{*}$-mode of $\nabla F$. In other
words, $\gamma$ adjusts itself according to $\beta$ in each iteration,
leading to $\beta$-independent algorithm. This is also verified via
simulations. 
\end{IEEEproof}
\begin{algorithm}
\begin{algorithmic}[1]

\Require{$\mathcal{T}$, $\beta$, $R,d$.}

\Ensure{$\mathbf{U}_{k},\boldsymbol{\Sigma}_{k},\mathbf{V}_{k},\forall k$:$\left\Vert \sum_{k=1}^{N}\left\llbracket \mathbf{U}_{k}\boldsymbol{\Sigma}_{k}\mathbf{V}_{k}^{T}\right\rrbracket _{\left(k,d\right)}\right\Vert _{TR}\leq\beta$.}

\State{Initialize $\mathcal{X}=0$, $R_{k}=r_{k}=0,\mathbf{U}_{k}=\Sigma_{k}=\mathbf{V}_{k}=\emptyset,\forall k$.}
\State{Initialize $\mathcal{N}=\left\{ 1,\dots,N\right\} .$}

\For{$n=1,2,\ldots,n_{\max}$}

\State{Set the tensor $\nabla F=\mathcal{X}(\mathcal{I})-\mathcal{T}(\mathcal{I})$.}

\State{Obtain $k^{*}=\arg\max_{k\in\mathcal{N}}\sigma_{max}\left[\left(\nabla F\right)_{\left(k,d\right)}\right]$.}

\State{Get the SVD $\left(\nabla F\right)_{\left(k^{*},d\right)}=\tilde{\mathbf{U}}_{k^{*}}\tilde{\Sigma}_{k^{*}}\tilde{\mathbf{V}}_{k^{*}}^{T}\in\mathbb{R}^{\bar{I}_{k^{*}}\times\bar{J}_{k^{*}}}$
}

\State{Update $r_{k}\leftarrow\min\left\{ \bar{I}_{k}-R_{k},\bar{J}_{k}-R_{k},R-\sum_{k=1}^{N}R_{k}\right\} $.}

\If{$r_{k}\leq0$}

\State{Break the loop.}

\EndIf

\State{Compute $\mathcal{S}$ from \eqref{eq:S_eqn}.}

\State{Get step size $\gamma$ from \eqref{eq:step_size}.}

\State{Update $\mathcal{X}\leftarrow\mathcal{X}-\gamma\mathcal{S}$.}

\State{Update $\mathbf{U}_{k^{*}},\boldsymbol{\Sigma}_{k^{*}},\mathbf{V}_{k^{*}}$.}

\State{Update $R_{k^{*}}=R_{k^{*}}+r_{k^{*}}$.}

\If{$R_{k^{*}}=\min\left\{ \bar{I}_{k^{*}},\bar{J}_{k^{*}}\right\} $}

\State{$\mathcal{N}\leftarrow\mathcal{N}\setminus\left\{ k^{*}\right\} $.}

\EndIf

\EndFor

\State{Return $\mathbf{U}_{k},\boldsymbol{\Sigma}_{k},\mathbf{V}_{k},\forall k$.}

\end{algorithmic}

\caption{Rank efficient modified FW algorithm, $TCA(\mathcal{T},R)$.\label{alg:ALGO1}}
\end{algorithm}

\subsubsection{Time and space complexity}

Let $\mathcal{X}$ be an $N$-order tensor of dimension $I\times\ldots\times I$.
In the algorithm \ref{alg:ALGO1}, rank-$R$ SVD is computed for the
cyclically unfolded matrix of size $I^{d}\times I^{N-d}$, which incurs
the complexity $\mathcal{O}\left(I^{2d}I^{N-d}\right)=\mathcal{O}\left(I^{N+d}\right)$.
Regarding space complexity for rank-$R$ decomposition, $I^{d}R+I^{N-d}R+R$
real valued space is required for $\mathbf{U}_{k},\mathbf{V}_{k},\Sigma_{k},\forall k$,
and $\left|\mathcal{I}\right|$ space for the observable tensor.

\subsection{Online prediction and caching algorithm}

Since the above tensor completion is fast, it can be used in an online
manner for edge caching, as shown in the Algorithm \ref{alg:ALGO2}.
In this algorithm, the tensor completion and linear prediction methods
are employed in the information exchange phase, whereas the cache
placement phase is dedicated to placing the content according to the
predicted number of normalized requests. 
\begin{algorithm}
\begin{algorithmic}[1]

\Require{$\mathcal{D}_{t},\forall t$, $R$.}

\Ensure{$\mathbf{c}_{bt},\forall b,t$}

\For{$t=\tau,\tau+1,\ldots$}

\State{\emph{IE phase}: Observe the tensor $\mathcal{D}_{t}$.}

\State{Apply $\mathcal{X}\leftarrow TCA(\mathcal{D}_{t},R)$.}

\State{Employ $\left\{ \hat{\bar{D}}_{fb,t+1},\forall f,b\right\} \leftarrow LP(\mathcal{X})$.
}

\State{\emph{CPL phase}: Based on $\hat{\bar{D}}_{fb,t+1},\forall f$,
obtain MPC placement $\mathbf{c}_{bt}$, for each $b$.}

\State{\emph{CD phase}: deliver contents from as per users' requests.
}

\EndFor

\end{algorithmic}

\caption{Online prediction and caching algorithm.\label{alg:ALGO2}}
\end{algorithm}

\section{Simulation Results\label{sec:Simulation-Results}}

Simulations are performed on the MovieLens dataset \cite{Harper:2015:MDH:2866565.2827872},
where fourth order tensors ($N=4$) are constructed from the movie
ratings with dimensions $F\times F\times N_{BS}\times\tau$, where
$F=128$, $N_{BS}=3$, and $\tau=10$. Each time slot entry is set
by aggregating the rating for 30 days based on the timestamps given.
$M=6$-th order prediction is performed. For tensor completion, values
$d=1$, $\beta=10^{5}$, $R=nN$, $n=2,4,6,8,10,12$ are chosen. For
edge caching, $L_{BS}=32$ is chosen. The algorithm is compared for
normalized reconstruction errors, defined as 
\begin{equation}
RSE=\frac{\left\Vert \mathcal{X}(\mathcal{I})-\mathcal{T}(\mathcal{I})\right\Vert _{F}}{\left\Vert \mathcal{T}(\mathcal{I})\right\Vert _{F}},
\end{equation}
which equals to $1$ at the start of first iteration, since $\mathcal{X}=0$.
. Two prediction methods are chosen, that is, linear predictions with
optimum coefficients (LP) and with equal coefficients (MP). The performance
of edge caching is measured in terms of cache hit rate. Algorithms
are run on a windows PC with Intel Xeon CPU E3-1230 v5 (3.40GHz, 32GB
RAM). 

\subsection{Convergence}

\begin{figure}
\centering\includegraphics[width=1\columnwidth]{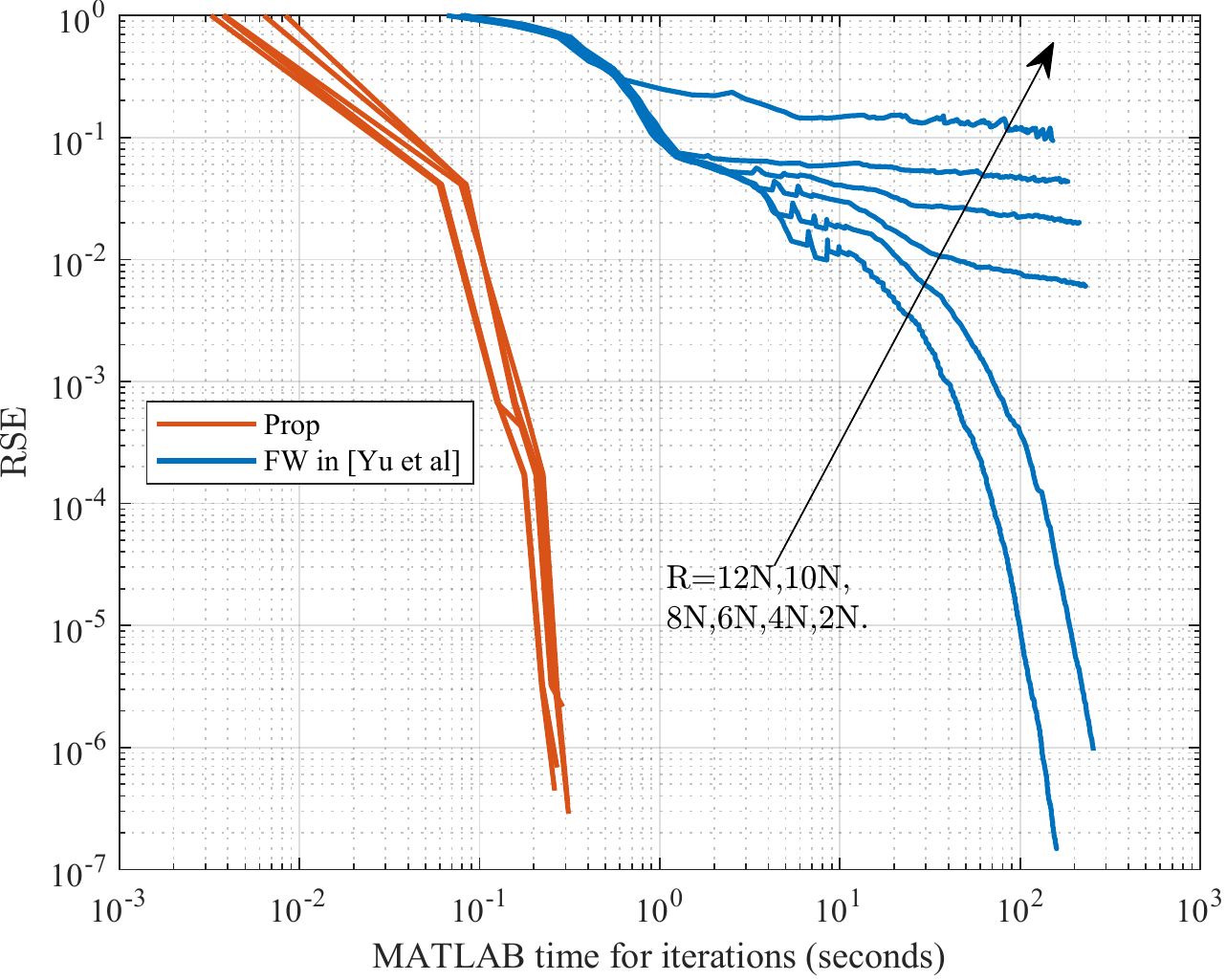}

\caption{RSE versus the execution time (in MATLAB) for the proposed algorithm
and the algorithm in \cite{9136661}, for $\beta=10^{5}$ and $R=nN$,
$\forall n=2,4,6,8,10,12$. \label{fig:Convergence}}
\end{figure}

Figure \ref{fig:Convergence} plots the convergence in terms of RSE
for the proposed algorithm, and the FW algorithm in \cite{9136661}
for $\beta=10^{5}$ and $R=nN$, $\forall n=2,4,6,8,10,12$. It can
be observed that the proposed algorithm outperforms significantly
as compared to the one in \cite{9136661}. It achieves less RSE in
less iterations (e.g. at $R=2N=32$, $RSE\approx10^{-1}$ for Yu \emph{et
al}, and $RSE<10^{-6}$ for the proposed one), and each iteration
executes in less time as well (total time is 0.2 seconds versus 200
seconds approximately). 

\subsection{Effect of latent factors $R$ and $\beta$}

\begin{figure}
\centering\includegraphics[width=1\columnwidth]{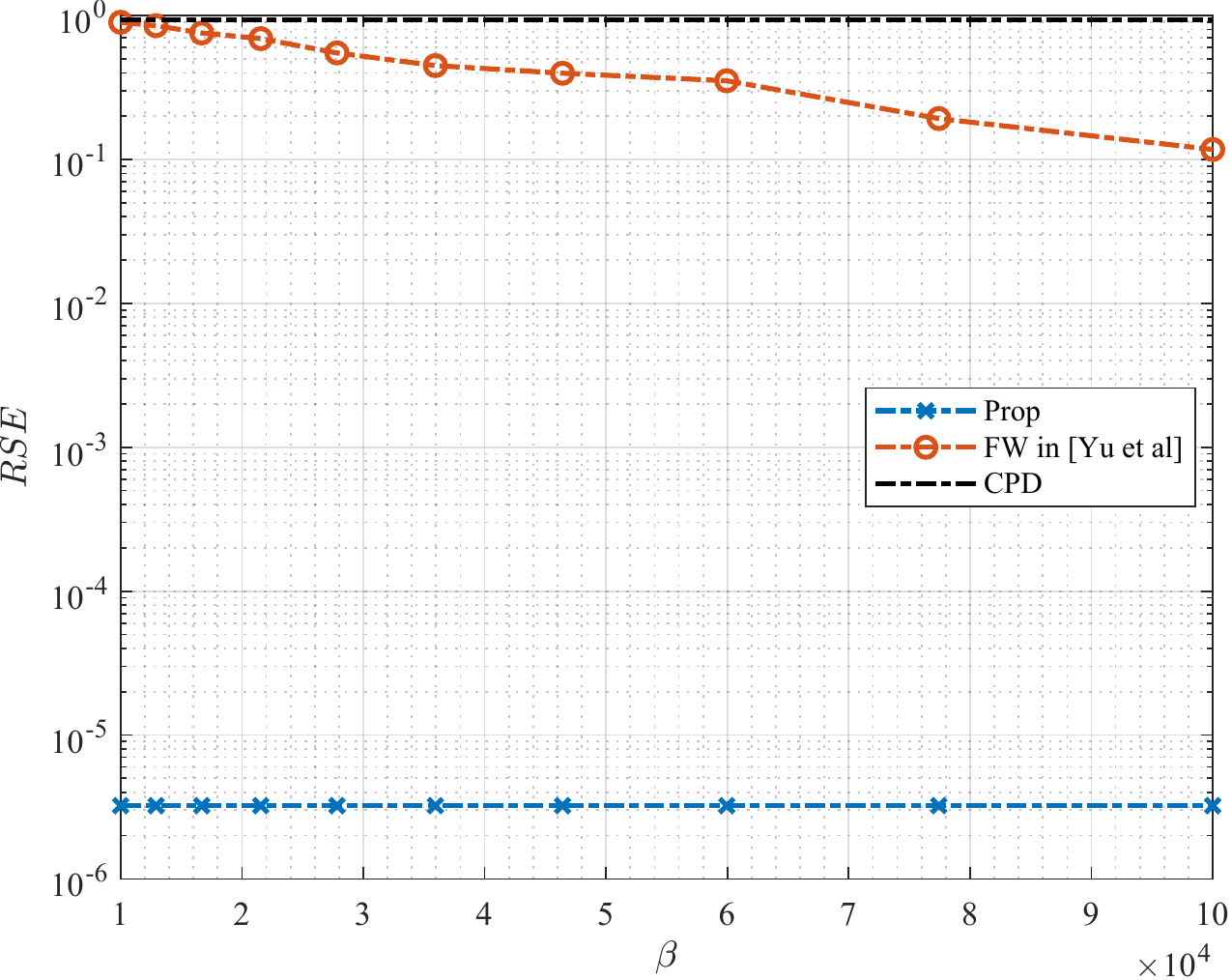}

\caption{Relative and reconstruction errors with respect to\textbf{ $\beta$}
at $R=6N$. \label{fig:BETA1}}
\end{figure}

\begin{figure}
\centering\includegraphics[width=1\columnwidth]{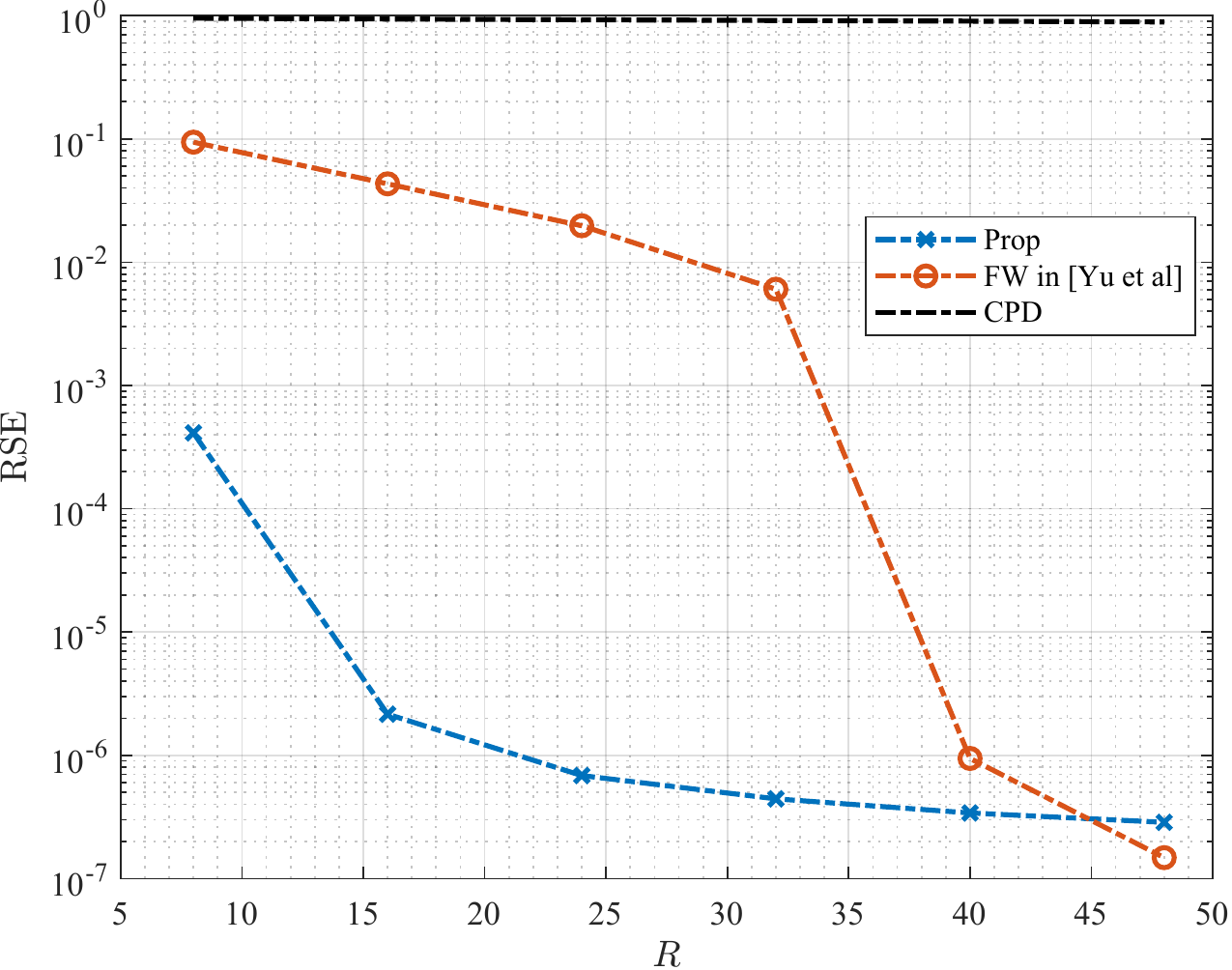}

\caption{Relative and reconstruction errors versus different ranks with fixed
$\beta$. \label{fig:RANK-plot}}
\end{figure}

To observe the effect of the factor $\beta$, Figure \ref{fig:BETA1}
plots the RSE with respect to $\beta$ at $R=6N$ for all the three
methods, while Figure \ref{fig:RANK-plot} shows with respect to the
rank $R$ with the fixed $\beta=10^{5}$. It can be seen that for
all values of $\beta$, the proposed algorithm provides much less
RSE than for CPD and the method in \cite{9136661}. Moreover, the
proposed algorithm is invariant to changes in $\beta$, as shown in
Proposition \ref{prop:The-propositionbeta}. Regarding the plots versus
rank-$R$, the proposed method performs significantly better at lower
ranks. As the rank is increased, the performance difference between
the proposed and the FW algorithm decreases. For sufficiently high
rank $R>45$, FW provides minutely better performance, where the difference
in RSE is the order of $10^{-6}$. 

\subsection{Cache hit rate }

\begin{figure}
\centering\includegraphics[width=1\columnwidth]{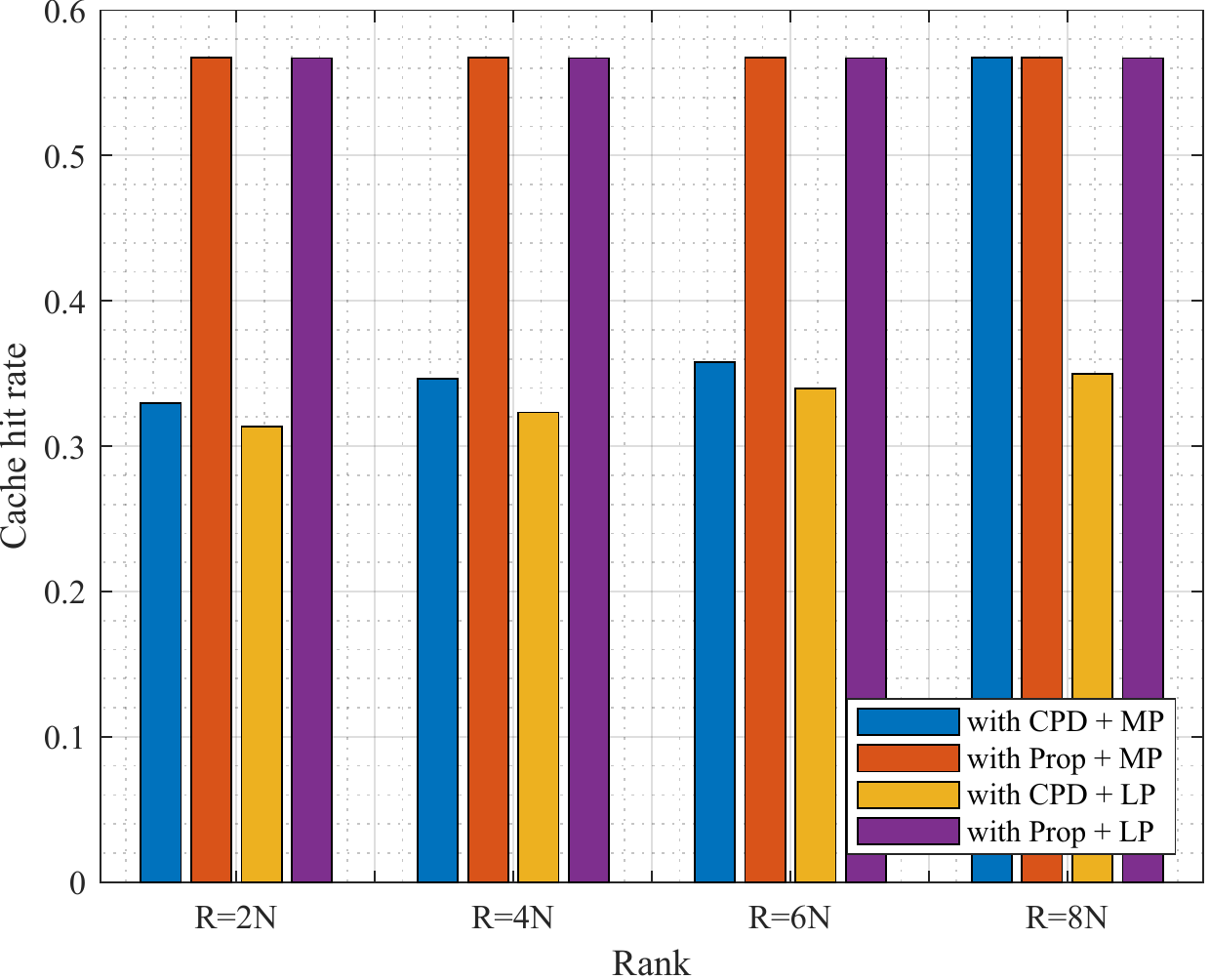}

\caption{Average cache hit rate for two linear prediction methods with three
different completion methods for MovieLens dataset with $R=2N,4N,6N$
ranks and $\beta=10^{5}$.\label{fig:CHR}}
\end{figure}

To evaluate the performance of the proposed method for edge caching
scenario, Figure \ref{fig:CHR} shows the normalized cache hit rates
averaged across time slots for three different methods with the cache
size of $32$. The results are averaged over 220 time slots, wherein
each time slot a tensor completion problem is solved and future popularity
is predicted using linear prediction (LP) and mean-prediction (MP)
methods. It can be observed that as compared to CPD, the CHR improvement
is around 75.8\%. Mean prediction and linear prediction perform approximately
similar for CHR. As the rank of completion is increased, improvements
in the CHR remains similar, which is due to rank-efficient tensor
completion. In other words, the proposed low-rank completion is efficient
at low ranks; and so, the completion at higher ranks yields similar
CHR performance, which can also be concluded from Figure \ref{fig:RANK-plot}.

\section{Conclusion\label{sec:Conclusion}}

In this paper, we have proposed an improved time complexity based
modified FW algorithm based on gradient descent, which has been shown
to significantly outperform in term of computational complexity and
performance. This algorithm needs only a few iterations, which is
of the order of the specified rank $R$. Further, for the wireless
edge caching, we have formulated content recommendation and prediction
problem into a tensor completion framework. Using the completed tensor,
we have employed linear prediction methods to obtain future popularities.
For this application, it is shown via simulations that after completion,
the algorithm provides significantly better cache hit rate (75\%)
as compared to the CP decomposition. 

For the edge caching, the library size is typically large of the order
of $10^{4}$. Performing tensor completion on such a large tensor
is both time and space intensive. Therefore, in future, we shall explore
to find the independent blocks in the observed tensor. It will significantly
reduce the time and space requirements and better tensor decomposition
can be found for the remaining block tensors as compared to the full
tensor. 

\bibliographystyle{IEEEtran}
\bibliography{tensor2,tensor1,caching_learning,caching_pred1,ppp1,caching1}

\end{document}